\documentclass[sigconf]{acmart}

\usepackage{multirow}
\AtBeginDocument{%
  \providecommand\BibTeX{{%
    \normalfont B\kern-0.5em{\scshape i\kern-0.25em b}\kern-0.8em\TeX}}}

\setcopyright{acmcopyright}
\copyrightyear{2022}
\acmYear{2022}
\acmDOI{10.1145/3503161.3548194}

\acmConference[MM '22] {Proceedings of the 30th ACM International Conference on Multimedia }{October 10--14, 2022}{Lisbon, Portugal.}
\acmBooktitle{Proceedings of the 30th ACM International Conference on Multimedia (MM '22), October 10--14, 2022, Lisbon, Portugal}
\acmPrice{15.00}
\acmISBN{978-1-4503-9203-7/22/10}

\acmSubmissionID{mmfp1854}

\newcommand{\tabincell}[2]{\begin{tabular}{@{}#1@{}}#2\end{tabular}}%

\begin{document}

\title{FastLTS: Non-Autoregressive End-to-End Unconstrained Lip-to-Speech Synthesis}

\author{Yongqi Wang}
\affiliation{%
  Zhejiang University
   \city{Hangzhou}
   \country{China}
}
\email{cyanbox@zju.edu.cn}

\author{Zhou Zhao}
\authornote{Corresponding author.}
\affiliation{%
  Zhejiang University
   \city{Hangzhou}
   \country{China}
}
\email{zhaozhou@zju.edu.cn}

\renewcommand{\shortauthors}{Wang, et al.}

\begin{abstract}
  Unconstrained lip-to-speech synthesis aims to generate corresponding speeches from silent videos of talking faces with no restriction on head poses or vocabulary. Current works mainly use sequence-to-sequence models to solve this problem, either in an autoregressive architecture or a flow-based non-autoregressive architecture. However, these models suffer from several drawbacks: 1) Instead of directly generating audios, they use a two-stage pipeline that first generates mel-spectrograms and then reconstructs audios from the spectrograms. This causes cumbersome deployment and degradation of speech quality due to error propagation; 2) The audio reconstruction algorithm used by these models limits the inference speed and audio quality, while neural vocoders are not available for these models since their output spectrograms are not accurate enough; 3) The autoregressive model suffers from high inference latency, while the flow-based model has high memory occupancy: neither of them is efficient enough in both time and memory usage.
  To tackle these problems, we propose FastLTS, a non-autoregressive end-to-end model which can directly synthesize high-quality speech audios from unconstrained talking videos with low latency, and has a relatively small model size. Besides, different from the widely used 3D-CNN visual frontend for lip movement encoding, we for the first time propose a transformer-based visual frontend for this task.
  Experiments show that our model achieves $19.76\times$ speedup for audio waveform generation compared with the current autoregressive model on input sequences of 3 seconds, and obtains superior audio quality.
\end{abstract}




\maketitle
\pagestyle{empty}

\begin{figure}[t]
    \includegraphics[width=\linewidth]{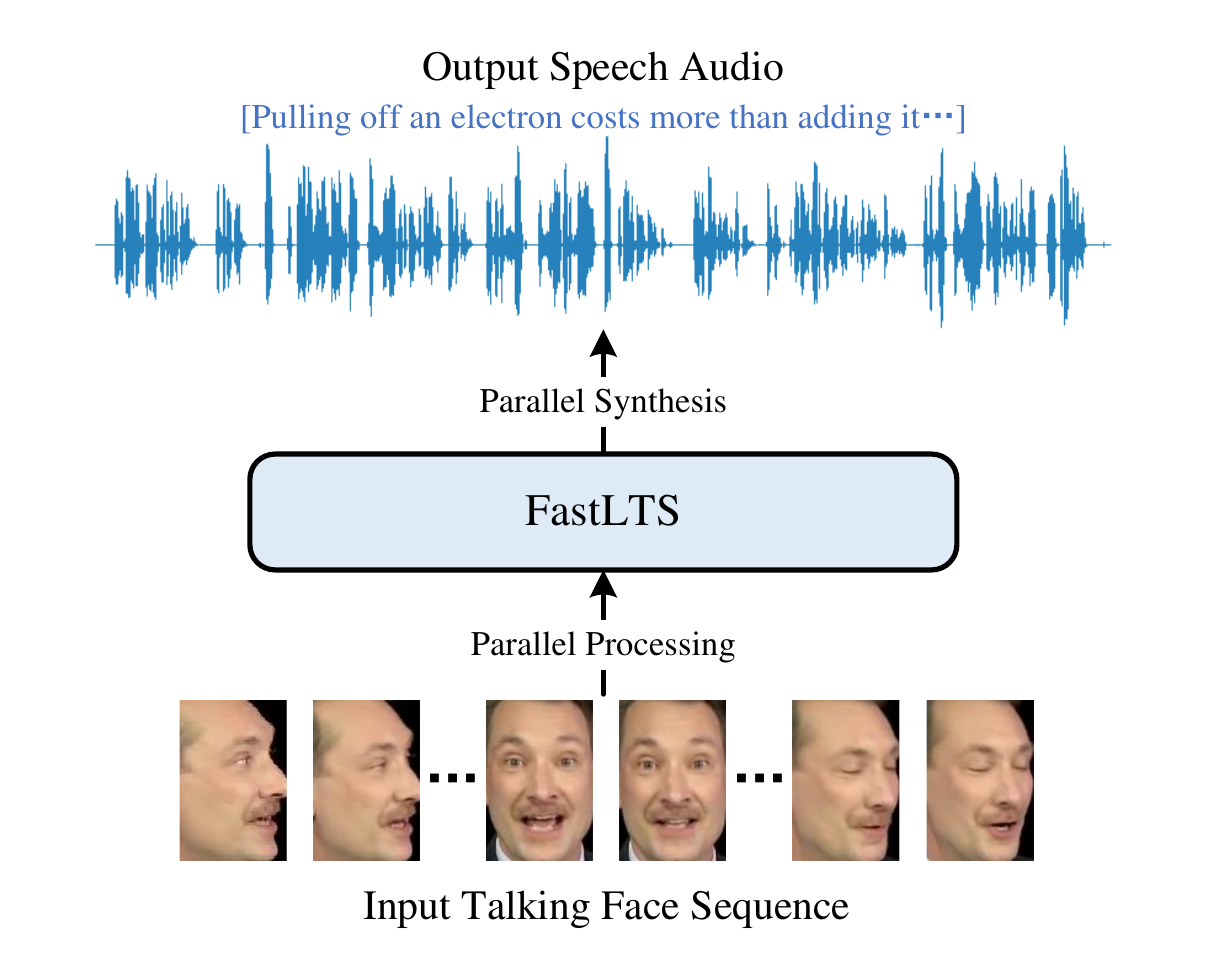}
    \caption{Illustration of end-to-end lip-to-speech synthesis. 
    Corresponding speech audios are generated conditioned on silent talking video frames. 
    Parallel generation is adopted in our non-autoregressive architecture.
    }
    \label{fig:teaser}
    \Description{teaser}
\end{figure}

\section{Introduction}

Lip-to-speech synthesis aims to reconstruct corresponding speech audios from 
silent videos of talking human faces by analyzing the facial movements, especially 
the movements of lips. This technology can help people understand speech in situations 
where voices are hardly audible, such as video conferencing in silent or noisy 
environment \cite{DBLP:conf/interspeech/Vougioukas0PP19}, long-range listening for surveillance \cite{DBLP:conf/iccvw/EphratHP17} and so on. Therefore, it has attracted a
big attention from researchers.

Most existing methods of lip-to-speech synthesis \cite{DBLP:conf/icassp/EphratP17,DBLP:conf/iccvw/EphratHP17,DBLP:conf/icassp/AkbariACM18, DBLP:conf/interspeech/Vougioukas0PP19, DBLP:conf/interspeech/MichelsantiSHGT20, yadav2021speech,DBLP:journals/corr/abs-2104-13332,kim2021lip} focus on constrained settings with small datasets, such as GRID \cite{cooke2006audio} and TCD-TIMIT \cite{7050271}, where each speaker's vocabulary contains less than a hundred words, and the speakers are always facing forward with almost no head movement. On the other hand, unconstrained lip-to-speech synthesis uses real-world talking videos which contain vocabulary of thousands of words and large head movements. This is much more difficult than the constrained settings, and most models for constrained lip-to-speech synthesis fail under the unconstrained settings \cite{DBLP:conf/cvpr/PrajwalMNJ20}. 

Current works \cite{DBLP:conf/cvpr/PrajwalMNJ20, GlowLTS} show that the sequence-to-sequence \cite{sutskever2014sequence,DBLP:conf/ssst/ChoMBB14} architecture is an effective solution to this problem. These works combine a visual encoder consisting of a stack of 3D-CNNs \cite{tran2015learning} and an LSTM \cite{schuster1997bidirectional} with acoustic models from TTS models \cite{DBLP:conf/icassp/ShenPWSJYCZWRSA18,DBLP:conf/nips/RenRTQZZL19,DBLP:conf/iclr/0006H0QZZL21}, either adopting an autoregressive architecture \cite{DBLP:conf/cvpr/PrajwalMNJ20} or a flow-based non-autoregressive architecture \cite{GlowLTS}. They use a two-stage pipeline that first generates mel-spectrograms as intermediate representations, and then synthesizes audio waveforms from the spectrograms with a signal-processing-based algorithm (i.e. Griffin-Lim algorithm \cite{1164317}). These works outperform previous works by a wide margin under unconstrained settings \cite{DBLP:conf/cvpr/PrajwalMNJ20, GlowLTS}.

Despite the impressive results achieved by these models, their two-stage pipeline design has some notable drawbacks. Such a pipeline increases the deployment cost of the model \cite{tan2021survey, DBLP:conf/icassp/WeissSBMK21}, and limits the audio quality due to error propagation and information loss in the intermediate representations \cite{kim2021conditional, DBLP:conf/iclr/DonahueDBES21}. Besides, the audio synthesis algorithm used in the second stage further limits the inference speed and audio quality, as it contains a large amount of serial calculations \cite{1164317}, and tends to produce low-quality audios with characteristic artifacts \cite{DBLP:conf/icassp/ShenPWSJYCZWRSA18}. Neural vocoders that are widely used in TTS models to synthesis high-quality audios, however, perform poor on these models, since their output spectrograms are not accurate enough for the vocoders to perform well \cite{DBLP:conf/cvpr/PrajwalMNJ20}.

In addition, both the autoregressive model \cite{DBLP:conf/cvpr/PrajwalMNJ20} and the flow-based non-autoregressive model \cite{GlowLTS} suffer from inefficiencies in either inference time or memory usage due to their model characteristics. The autoregressive model suffers from high inference latency due to its recursive nature\cite{DBLP:conf/iclr/Gu0XLS18,guo2019non,DBLP:conf/nips/RenRTQZZL19}, while the flow-based model has a large amount of parameters which causes high memory occupancy (see Section 5.6).
These drawbacks significantly limit the performance of these models.

In this paper, we propose FastLTS, a non-autoregressive end-to-end model for unconstrained lip-to-speech synthesis. Our model is able to directly synthesize high-quality speeches with much less latency. Specifically, we use a transformer-based non-autoregressive acoustic decoder and a GAN-based vocoder to realize a fully parallelized end-to-end model. We show that our model significantly reduces inference time while improving the audio quality, and has a relatively light model size.

Moreover, transformer architecture has recently achieved equivalent or superior performance to convolutional networks on image and video classification \cite{DBLP:conf/iclr/DosovitskiyB0WZ21,DBLP:conf/iccv/Arnab0H0LS21,DBLP:conf/icml/BertasiusWT21}, proving its ability of visual feature extraction. However, most lipreading and lip-to-speech synthesis models still use 3D-CNNs for visual feature extraction. A recent work \cite{DBLP:journals/corr/abs-2201-10439} shows that a ViT\cite{DBLP:conf/iclr/DosovitskiyB0WZ21}-based visual frontend outperforms 3D-CNN counterparts in audio-visual automatic speech recognition. However, original ViT calculates self-attention among all tokens from a video, causing huge computation and memory burden, thus is inappropriate for long video sequences.
 
 In this work, we devise a transformer-based visual frontend for visual feature extraction in lip-to-speech synthesis. Experiments show that it has viable feature extracting ability on both large unconstrained datasets and small constrained datasets. As far as we know, this is the first practical transformer-based visual frontend for lip-to-speech synthesis and lipreading.
 
 To conclude, our key contributions are as follows:
 \begin{itemize}
     \item We propose an end-to-end model for unconstrained lip-to-speech synthesis, which is able to synthesize high-quality speeches with much less inference latency.
     \item We propose a transformer-based visual frontend for lip movement encoding, which shows good performance on both large and small datasets.
     \item Experiments show that our model achieves $9.14\times$ speedup for mel-spectrogram generation and $19.76\times$ speedup for audio waveform generation compared with the autoregressive model on input sequences of 3 seconds, and obtains superior audio quality.
 \end{itemize}
 
\section{Related Works}
\subsection{Constrained Lip-to-Speech Synthesis}

Constrained lip-to-speech synthesis aims to generate speech audios from silent talking videos where the vocabulary is narrow and the speakers' heads have almost no movement. Ephrat et al. \cite{DBLP:conf/icassp/EphratP17} propose the first solution to this problem, which uses CNN to predict LPC (Linear Predictive Coding) features from silent talking videos. Later they augment their model to a two-tower CNN-based encoder-decoder model \cite{DBLP:conf/iccvw/EphratHP17} and encode raw frames and optical flows respectively. Akbari et al. \cite{DBLP:conf/icassp/AkbariACM18} combine an autoencoder that extracts bottleneck features from audio spectrograms with a lipreading network for visual feature extracting. Yadav et al. \cite{yadav2021speech} use stochastic modeling approach with a variational autoencoder. Vougiouskas et al. \cite{DBLP:conf/interspeech/Vougioukas0PP19} and Mira et al. \cite{DBLP:journals/corr/abs-2104-13332} use GAN to directly synthesize audio waveform from video frames. Kim et al. \cite{kim2021lip} combine GAN with attention mechanism for generating better mel-spectrograms.

Although these works show impressive results on small constrained datasets, they fail to handle unconstrained datasets with large head movements and large vocabulary, which limits their applications in real-world scenarios.
Different from these models, our model mainly focuses on unconstrained lip-to-speech synthesis. Yet we also conduct experiments on GRID \cite{cooke2006audio} dataset to prove the modeling ability on small constrained datasets of our model.

\begin{figure*}
    \includegraphics[width=\textwidth]{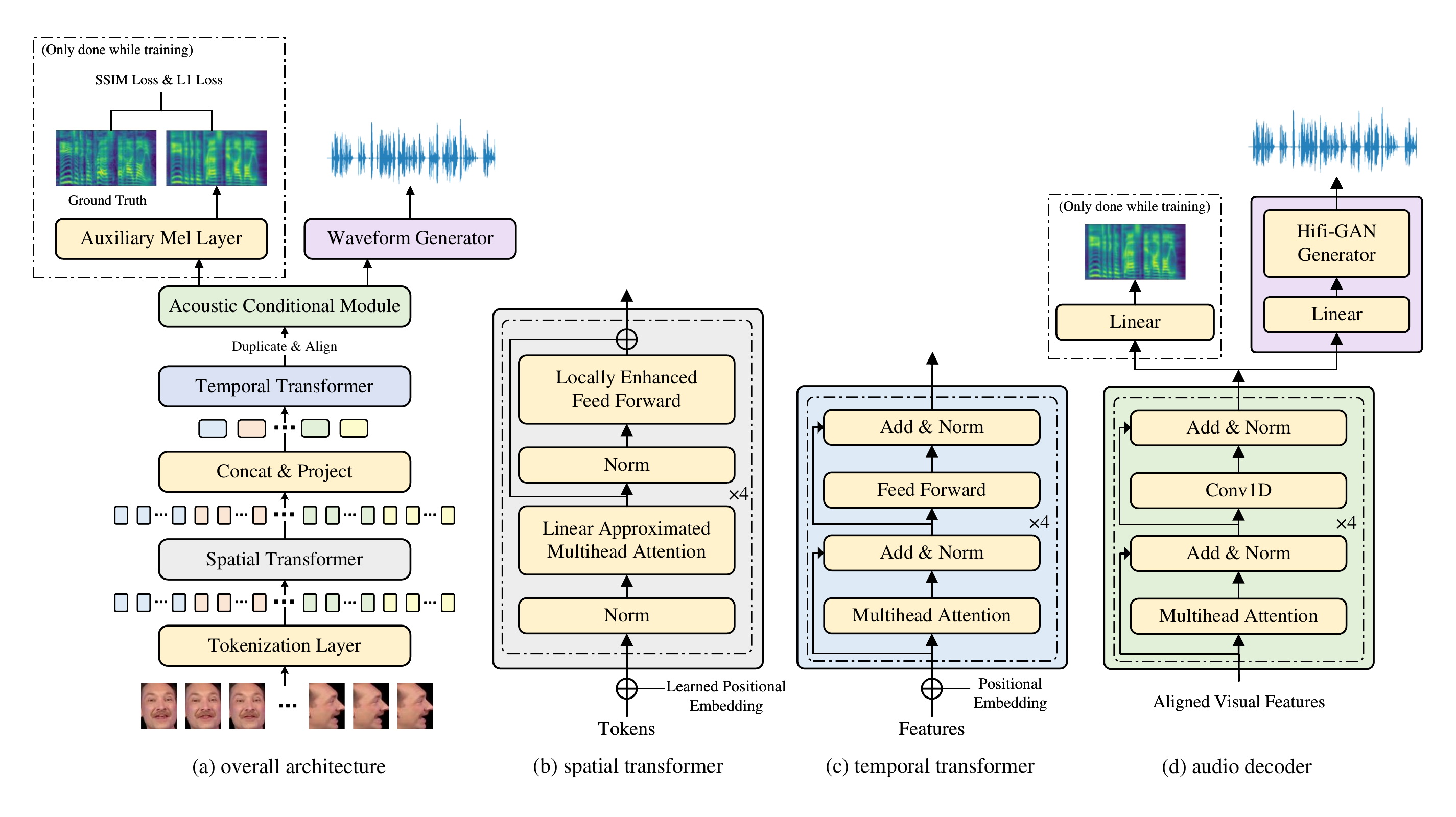}
    \caption{(a) The overall architecture of the model (b) The spatial transformer block (c) The temporal transformer block (d) The audio decoder including the acoustic conditional module and the waveform generator.}
    \label{fig:overall}
    \Description{architecture}
\end{figure*}

\subsection{Unconstrained Lip-to-Speech Synthesis}
Unconstrained lip-to-speech synthesis aims to generate speeches from real-world talking videos with large vocabulary and head movements. As this topic has only recently attracted attention of the researchers, there are not many works on it currently. Prajwal et al. \cite{DBLP:conf/cvpr/PrajwalMNJ20} firstly propose an autoregressive sequence-to-sequence model modified from Tacotron 2 \cite{DBLP:conf/icassp/ShenPWSJYCZWRSA18} to tackle this problem, which generates mel-spectrograms conditioned on video frames; He et al. \cite{GlowLTS} use a non-autoregressive architecture to accelerate inference and use a Glow \cite{DBLP:conf/nips/KingmaD18} module for mel-spectrogram refinement.

While these models find feasible solution to unconstrained lip-to-speech synthesis, their two-stage pipeline using Griffin-Lim algorithm limits the audio quality and inference speed. Besides, they suffer from low temporal or spatial efficiency due to their model characteristics. Different from these models, our model uses an end-to-end non-autoregressive architecture, which is efficient on both inference time and model size, and also improves the audio quality.

\subsection{Vision Transformers}
Transformer \cite{DBLP:conf/nips/VaswaniSPUJGKP17} has shown decent performance in multiple NLP tasks since it was proposed. Dosovitskiy et al. \cite{DBLP:conf/iclr/DosovitskiyB0WZ21} bring transformer to the area of computer vision for the first time, and demonstrate that transformer architecture has comparable or even superior performance to convolutional networks on image classification. Later, transformer architecture is extended to videos. Arnab et al. \cite{DBLP:conf/iccv/Arnab0H0LS21} propose a spatial-temporal factorized transformer encoder architecture for video encoding. Several other works \cite{neimark2021video,DBLP:conf/icml/BertasiusWT21,sharir2021image} investigate different attention mechanism for video processing.
On the other hand, some works add convolution mechanism to transformers, combining the advantages of convolution and attention mechanism. Wu et al. \cite{wu2021cvt} use convolutional networks to calculate self-attention. Yuan et al. \cite{yuan2021incorporating} add convolution to tokenization layer and feed forward network. Inspired by these works, we propose a spatial-temporal factorized transformer visual encoder for lip-to-speech synthesis. Experiments show that the proposed frontend has comparable encoding ability to traditional convolution networks.

\section{Problem Definition}

In this section, we introduce the problem formulation of unconstrained lip-to-speech synthesis. Suppose we have a video sequence of a single talking human $V=\{v_1, v_2, ..., v_T\}$, where $T$ denotes the length of the video sequence and $v_i$ denotes the \textit{i}-th frame. The task of lip-to-speech synthesis is to generate the corresponding speech audio $A=\{a_1, a_2, ..., a_L\}$, where $L$ denotes the length of the generated audio, and $a_j$ denotes the $j$-th value of the waveform. The relationship between $T$ and $L$ is
\begin{equation}
    T \cdot {sr} = L \cdot {FPS}
\end{equation}
where $sr$ is the sampling rate of the audio and ${FPS}$ is the frame rate of the video.

The word "unconstrained" means that the head positions of two frames $v_i$ and $v_j$ may have a big difference where $i$ and $j$ are not very close, and the corresponding speech can have a vocabulary of thousands of words. Constrained lip-to-speech synthesis can be seen as a special case of the unconstrained one.

\section{Methods}

In this section, we illustrate the architecture of FastLTS and introduce our training methods. As shown in Figure \ref{fig:overall}(a), the model mainly consists of three parts: the visual encoder (from the tokenization layer to the temporal transformer), the acoustic conditional module, and the waveform generator. A linear layer is used for auxiliary mel-spectrogram output while training. The input video frames are fed into the tokenization layer, then the features are feed forward through the whole model, and finally taken by the waveform generator to synthesize waveforms. We will introduce these modules in detail in the following subsections.

\subsection{Visual Encoder}
The visual encoder is used to extract and encode visual features from the input video sequence. It consists of a tokenization layer, a spatial transformer and a temporal transformer. The tokenization layer is used to preliminarily extract local features and produce spatio-temporal tokens for the transformer. It consists of a 3D-convolutional layer, a layer normalization layer and a max-pooling layer. The tokens are added with a learned positional embedding and are then fed into the spatial transformer, the architecture of which is illustrated in Figure \ref{fig:overall}(b). The spatial transformer is used to model the correlation among spatially adjacent tokens, and only calculates attentions on tokens extracted from the same temporal index. However, since the tokens are overlapped while doing tokenization, tokens of the same temporal position may still form a relatively long sequence, causing large computation burden and memory usage. To deal with this problem, we use linear approximation of self-attention proposed in Performer \cite{DBLP:conf/iclr/ChoromanskiLDSG21}, which significantly reduces the computation burden of self-attention while keeping a good performance. For the feed forward part of the spatial transformer, we employ the Locally Enhanced Feed Forward network proposed in \cite{yuan2021incorporating}, which does convolution on the depth dimension of the features. This brings convolution into the transformer architecture, which strengthens its locality and gives it stronger ability to model spatial correlation among neighboring tokens.

The output hiddens of the spatial transformer with the same temporal index are then concatenated and projected to a single hidden of lower dimension by a linear layer. This sequence is then fed into the temporal transformer, which is illustrated in Figure \ref{fig:overall}(c). The temporal transformer models the temporal correlation between the hiddens. Its structure is the same as the original transformer proposed in \cite{DBLP:conf/nips/VaswaniSPUJGKP17}.

\subsection{Acoustic Conditional Module}
We notice that it is very difficult to directly use audio waveforms as supervisory signal to train the model due to its high dimensionality. Therefore, we employ an acoustic conditional module, which is illustrated in the green part of Figure \ref{fig:overall}(d). This module works as a non-autoregressive decoder that turns visual features into acoustic features. Basically it is a transformer which alternates the feed forward network with a 1D convolutional network for better locality modeling of the acoustic features. It can output mel-spectrograms when combined with the auxiliary mel layer. Note that this module does not feed mel-spectrograms into the waveform generator, and the auxiliary mel layer is only used to assist training. We will explain the training methods in detail later.

Note that before the output visual features of the temporal transformer are fed into the acoustic conditional module, they need to be aligned to match the length of the acoustic feature sequence. We simply duplicate the visual features for alignment. We calculate the duplication factor $d$ with the formula $d=L_{aud}/FPS$, where $L_{aud}$ is the length of the audio feature sequence per second. If $d$ is an integer, we simply duplicate the visual features by $d$ times. Otherwise, we divide the visual features into two parts and duplicate them by $\lfloor d \rfloor$ and $\lceil d \rceil$ times respectively, and adjust the proportion of the two parts for alignment. For example, if the frame rate is 30 fps and the length of the audio features is 80 per second, we use duplication factors of $\{3, 3, 2, 3, 3, 2, ...\}$.

\subsection{Waveform Generator}

The structure of the waveform generator is illustrated in the violet part of Figure \ref{fig:overall}(d). The linear layer has the same dimension as the auxiliary mel layer, and it projects the acoustic features to the same dimension as the mel-spectrogram frame, which is 80 in our case. We then feed the projected features to the generator to synthesize audio waveform. The structure of the generator is the same as that proposed in \cite{DBLP:conf/nips/KongKB20}. We refer the readers to the Hifi-GAN paper \cite{DBLP:conf/nips/KongKB20} for more details about the generator.

\subsection{Training Method}

As mentioned in section 4.2, due to the high dimensionality of the raw audio, it is very difficult to directly use it as supervisory signal to train the whole model in an end-to-end way. We therefore adopt a two-stage training method. In the first stage, we don't involve the waveform generator in training, and only train the visual encoder and the acoustic conditional module. We use the auxiliary mel layer to output mel-spectrograms and use SSIM loss \cite{DBLP:journals/tci/ZhaoGFK17} and L1 loss to optimize the encoder and the conditional module. The SSIM loss is formulated as:
\begin{equation}
    \mathcal{L}_{\rm{SSIM}} = \frac{1}{L_{mel}}\sum_{n=1}^{L_{mel}}1-{\rm SSIM}(y_n, \hat{y}_n)
\end{equation}
Where $L_{mel}$ denotes the length of the mel-spectorgram, $y_n$ denotes the $n$-th frame of ground truth mel-spectrogram, $\hat{y}_n$ denotes the $n$-th frame of generated mel-spectrogram, and $\rm SSIM(\cdot,\cdot)$ denotes the structural similarity index \cite{1284395} of two vectors. The L1 loss is formulated as:
\begin{equation}
    \mathcal{L}_{L1} = \frac{1}{L_{mel}} \sum_{n=1}^{L_{mel}} \lVert y_{n}  - \hat{y}_{n} \rVert_1
\end{equation}
And the total loss function of the first stage is:
\begin{equation}
    \mathcal{L}_{stage1} = \lambda_{\rm SSIM}\mathcal{L}_{{\rm SSIM}} + \lambda_{L1}\mathcal{L}_{L1}
\end{equation}

In the second stage, we remove the auxiliary mel-spectrogram output, and plug the waveform generator into the model. We use the same loss as \cite{DBLP:conf/nips/KongKB20} to train the whole model. Specifically, we use a multi-period discriminator and a multi-scale discriminator to extract feature maps of the audios and discriminate the audio samples for adversarial training. The training objective of the second stage consists of three parts: adversarial loss, mel-spectrogram loss and feature matching loss. The adversarial loss for discriminators $D$ and generator $G$ is:
\begin{eqnarray}
& \mathcal{L}_{adv}(D;G)=\mathbb{E}_{(x,s)}\bigg [(D(x)-1)^2+(D(G(s)))^2 \bigg ]
\\
& \mathcal{L}_{adv}(G;D)=\mathbb{E}_{(x,s)}\bigg [(D(G(s))-1)^2 \bigg]
\end{eqnarray}
Where $s$ is the input condition and $x$ is the ground truth audio.

The mel-spectrogram loss is defined as:
\begin{equation}
    \mathcal{L}_{mel}(G)=\mathbb{E}_{(x,s)}\bigg[\lVert \phi(x) -\phi(G(s)) \rVert_1\bigg]
\end{equation}
where $\phi(\cdot)$ denotes the function to transform a waveform into corresponding mel-spectrogram. And the feature matching loss is defined as:
\begin{equation}
\mathcal{L}_{F M}(G ; D)=\mathbb{E}_{(x, s)}\left[\sum_{i=1}^{T} \frac{1}{N_{i}}\left\|D^{i}(x)-D^{i}(G(s))\right\|_{1}\right]
\end{equation}
where $T$ denotes the number of layers in the discriminator; $D^i$ and $N^i$ denote the features and the
number of features in the $i$-th layer of the discriminator, respectively.

The total loss of the second stage is:
\begin{eqnarray}
  &  \mathcal{L}_{stage2-G} = \lambda_{a}\mathcal{L}_{adv}(G;D)+\lambda_{m}\mathcal{L}_{mel}(G)+\lambda_{f}\mathcal{L}_{FM}(G;D)
    \\
  &  \mathcal{L}_{stage2-D}=\mathcal{L}_{adv}(D;G)
\end{eqnarray}

We find that calculating loss on a whole segment of audio leads to large computation cost and memory usage due to the high resolution of audio. Therefore, we adopt the sampling window technique for wave generator training. Specifically, we sample a consecutive
sub-sequence from the acoustic conditional module output, and use the corresponding audio segment to train the waveform generator. We notice that for small dataset, this will reduce computation cost and memory occupancy while keeping a good performance. However, for large unconstrained dataset, training the whole model in this way will damage the encoding ability of the visual encoder due to the shrink of context window. Therefore, we freeze the parameters of the visual encoder and the acoustic conditional module in the second stage and only optimize the parameters of the waveform generator.

\section{Experiments and Results}

\subsection{Datasets}

\textbf{Lip2Wav} The Lip2Wav dataset \cite{DBLP:conf/cvpr/PrajwalMNJ20} is the largest and most commonly used dataset for unconstrained lip-to-speech synthesis. It contains real-world lecture videos of 5 different speakers, with about 20 hours of videos per speaker and vocabulary size over 5000 words for each of them. We conduct our experiments on 3 speakers, which are \textit{Chess Analysis}, \textit{Chemistry Lectures} and \textit{Hardware Security}. All three speakers has frame rate of 30 fps, and we sample the raw audios at 16000Hz. We use the same train-test splits as \cite{DBLP:conf/cvpr/PrajwalMNJ20} for training and evaluation.

\noindent\textbf{GRID} The GRID dataset \cite{cooke2006audio} is a symbolic dataset for constrained lip-to-speech synthesis. It contains 34 speakers with 1000 sentences uttered by each speaker. The talking videos are captured in an artificial environment. The vocabulary of GRID contains only 51 words. All the sentences are in a restricted grammar and each sentence contains 6 to 10 words. We conduct our experiments on three speakers in the GRID dataset to prove the adaptability on small constrained datasets of our model.

\subsection{Implementation Details}

\begin{figure*}
    \includegraphics[width=\textwidth]{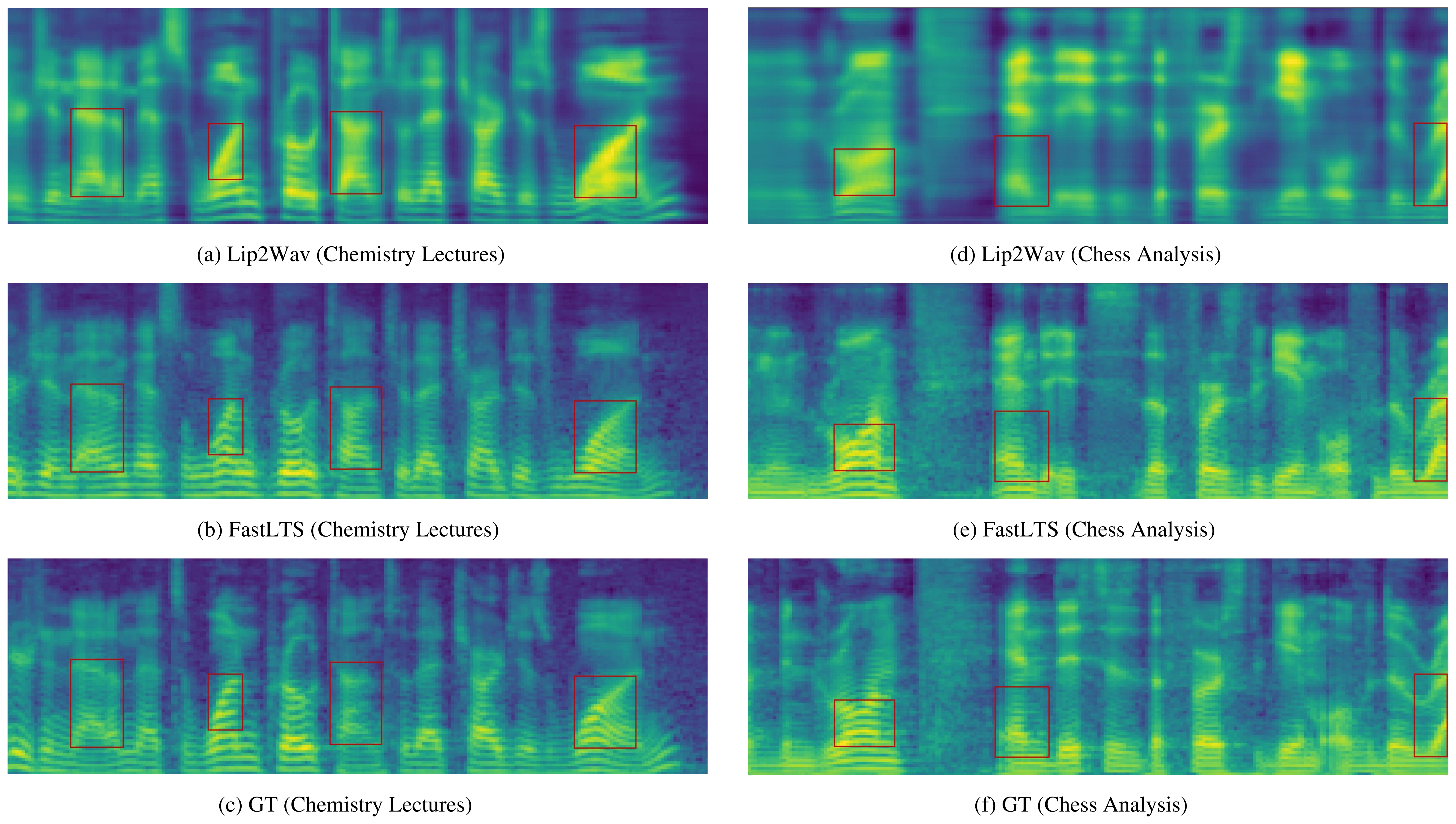}
    \caption{Comparison of mel-spectrograms from different methods}
    \label{fig:melcomp}
    \Description{architecture}
\end{figure*}

\textbf{Data Preprocessing.} We use 3-second contiguous sequences as training samples. For audios, we sample the audios at 16000Hz, and transform them to mel-spectrograms with window-size being 800, hop-size being 200 and spectrogram dimension being 80. For video frames, we use $S^3FD$ face detector to detect the faces and crop them out. We then resize the cropped faces to $96\times96$. We use data augmentation for the GRID dataset to improve performance and avoid overfitting. Specifically, we flip the facial images horizontally with probability of 40\%, and we randomly crop 0-7.2\% horizontal or vertical pixels of the images with probability of 40\%.

\noindent\textbf{Model Configurations.} The kernel size of the 3D-convolutional layer in the tokenization layer is set to $5\times5\times5$. The dimension of the visual tokens $d_{token}$ is set to 32. The spatial transformer is a 4-layer transformer with hidden dimension $d_{s}$ being 36, number of heads $h_{s}$ being 6. The temporal transformer is also a 4-layer transformer with number of heads $h_{t}$ being 8. We set the temporal hidden dimension $d_{t}$ to 384 for the Lip2Wav dataset, and 160 for the GRID dataset. The output dimension of the feed forward layer $d_{ff}$ is $4\times d_t$. The acoustic conditional module has the same dimension of hiddens and number of layers as the temporal transformer. For the waveform generator, the kernel sizes of the transposed convolutions $k_u$ and the resblocks $k_r$ are set to $[9,9,8,4]$ and $[3,7,11]$. The sampling window size is set to 1.2 seconds. For the two training stage, we use Adam and AdamW optimizers, and the initial learning rates are set to $2\times10^{-3}$ and $2\times10^{-4}$, respectively.

\begin{table}
  \caption{MOS on Lip2Wav Dataset}
  \label{tab:l2w_mos}
  \begin{tabular}{clccc}
    \toprule
    \textbf{Speaker} & \textbf{Method} & \textbf{Quality} & \textbf{Intelli.} & \textbf{Natural.} \\
    \midrule
     \multirow{3}{*}{\tabincell{c}{Chess\\Analysis}} & Lip2Wav & $3.53\pm 0.10$ & $3.51\pm 0.09$ & $3.48\pm 0.09$\\
      & FastLTS & $3.79\pm 0.09$ & $3.82\pm 0.10$ & $3.59\pm 0.08$ \\
     & GT & $4.08\pm 0.07$ & $3.91\pm 0.08$ & $4.10\pm 0.06$ \\
     \midrule
     \multirow{3}{*}{\tabincell{c}{Chemistry\\Lectures}} & Lip2Wav & $3.60\pm 0.10$ & $3.88\pm 0.10$ & $3.78\pm 0.09$\\
      & FastLTS & $3.84\pm 0.10$ & $3.73\pm 0.11$ & $3.87\pm 0.09$\\
     & GT & $4.06\pm 0.08$ & $3.90\pm 0.09$ & $4.10\pm 0.07$ \\
     \midrule
     \multirow{3}{*}{\tabincell{c}{Hardware\\Security}} & Lip2Wav & $3.67\pm 0.10$ & $3.57\pm 0.11$ & $3.74\pm 0.11$ \\
      & FastLTS & $3.86\pm 0.12$ & $3.89\pm 0.13$ & $3.80\pm 0.14$ \\
     & GT & $3.94\pm0.10$ & $4.01\pm0.10$ & $4.02\pm0.09$ \\
    
  \bottomrule
\end{tabular}
\end{table}

\begin{table}
  \caption{MOS of Lip2Wav + Vocoder ( Failed Cases )}
  \label{tab:l2wvoc_mos}
  \begin{tabular}{cccc}
    \toprule
    \textbf{Speaker} &  \textbf{Quality} & \textbf{Intelligibility} & \textbf{Naturalness} \\
    \midrule
     \tabincell{c}{Chess\\Analysis} & $1.35\pm 0.05$ & $1.81\pm 0.04$ & $1.54\pm 0.06$ \\
     \midrule
     \tabincell{c}{Chemistry\\Lectures} & $1.89\pm 0.04$ & $1.47\pm 0.06$ & $1.91\pm 0.03$ \\
  \bottomrule
\end{tabular}
\end{table}

\subsection{Evaluation Metrics}
We mainly evaluate the generated audios through subjective evaluation on the quality, intelligibility and naturalness of the audios. Here, quality indicates the clarity of audio against electronic noise and distortion, naturalness indicates how much the prosody and rhythm of the speech sound like real human, and intelligibility indicates the clarity of the pronunciation and the difficulty of understanding the speech content. Our subjective evaluation is conducted via Amazon Mechanical Turk. For each test item, we ask over 300 respondents to fill out three 1-5 Likert scales on 
the above three metrics of the generated and ground truth audios and report the mean opinion score (MOS). We paid the respondents around \$7.2 hourly. We also calculate PESQ \cite{941023} of our results on the GRID dataset for objective quality evaluation. For inference speed evaluation, we measure the inference speed of mel-spectrograms and waveforms of different models on input sequences of different lengths.

Different from previous works, we don't calculate STOI \cite{5713237} and ESTOI \cite{jensen2016algorithm} to measure the audio intelligibility, as we claim that these criteria are not appropriate for GAN-based models that directly synthesize audio waveforms. These algorithms measure the distortion of a noisy signal relative to the original one, while GANs may produce intelligible speeches with different intonation from the original speeches, which causes a relatively poor STOI value yet does no damage to intelligibility and naturalness. We therefore mainly use human evaluation for intelligibility evaluation. Later we will release demonstration videos to prove our results.

\subsection{Audio Quality}

For the Lip2Wav dataset, we conduct human evaluation on the results of FastLTS and Lip2Wav together with the ground truth audios, and present the MOS in Table \ref{tab:l2w_mos}, including mean scores and confidence intervals with confidence level of 95\%. We can see that our model achieves significant improvement on audio quality, and improves the speech intelligibility a lot on two speakers. This proves that through adopting an end-to-end architecture with a vocoder, our model is able to generate clearer audios with less noise and distortion, and makes the pronunciation more accurate. Our model also achieves higher score on naturalness, which indicates that it is able to generate more human-like voices with the assistance of adversarial training. The above experimental results also indicate that our transformer-based visual frontend has comparable performance to the 3D-CNN counterparts used in previous works.

We visualize the mel-spectrogram samples generated by different methods in Figure \ref{fig:melcomp} in order to show the superiority of our model more intuitively. We transform the outputs of FastLTS and ground truth audios to mel-spectrograms and compare them with the outputs of Lip2Wav. The left column shows results on speaker of \textit{Chemistry Lectures}, and the right column shows those on speaker of \textit{Chess Analysis}. We can see that the spectrograms generated by Lip2Wav tend to be blurry and over-smoothed, while the spectrograms from our model have clear and sharp edges. We marked some areas with red boxes, where the results of the two methods form a striking contrast.

\begin{table}
  \caption{MOS on GRID Dataset}
  \label{tab:grid_mos}
  \begin{tabular}{lccc}
    \toprule
    \textbf{Method} & \textbf{Quality} & \textbf{Intelligibility} & \textbf{Naturalness} \\
    \midrule
     Lip2Wav & $3.27\pm 0.11$ & $3.47\pm 0.13$ & $3.54\pm 0.12$\\
      FastLTS & $3.59\pm0.09$ & $3.68\pm0.09$ & $3.73\pm0.08$ \\
     GT & $3.60\pm0.10$ & $3.76\pm0.09$ & $3.78\pm0.10$ \\
  \bottomrule
\end{tabular}
\end{table}

\begin{table}
  \caption{PESQ on GRID dataset}
  \label{tab:gridpesq}
  \begin{tabular}{lc}
    \toprule
    \textbf{Method} & \textbf{PESQ} \\
    \midrule
    Vid2Speech \cite{DBLP:conf/icassp/EphratP17} & 1.734\\
    Lip2AudSpec \cite{DBLP:conf/icassp/AkbariACM18} & 1.673 \\
    GAN-based \cite{DBLP:conf/interspeech/Vougioukas0PP19} & 1.684 \\
    Ephrat et al. \cite{DBLP:conf/iccvw/EphratHP17} & 1.825 \\
    Lip2Wav \cite{DBLP:conf/cvpr/PrajwalMNJ20} & 1.772\\
    VAE-based \cite{yadav2021speech} & 1.932 \\
    Vocoder-based \cite{DBLP:conf/interspeech/MichelsantiSHGT20} & 1.900 \\
    VCA-GAN \cite{kim2021lip} & \textbf{2.008} \\
    \midrule
    FastLTS & 1.939 \\
  \bottomrule
\end{tabular}
\end{table}

\begin{figure*}
    \includegraphics[width=\textwidth]{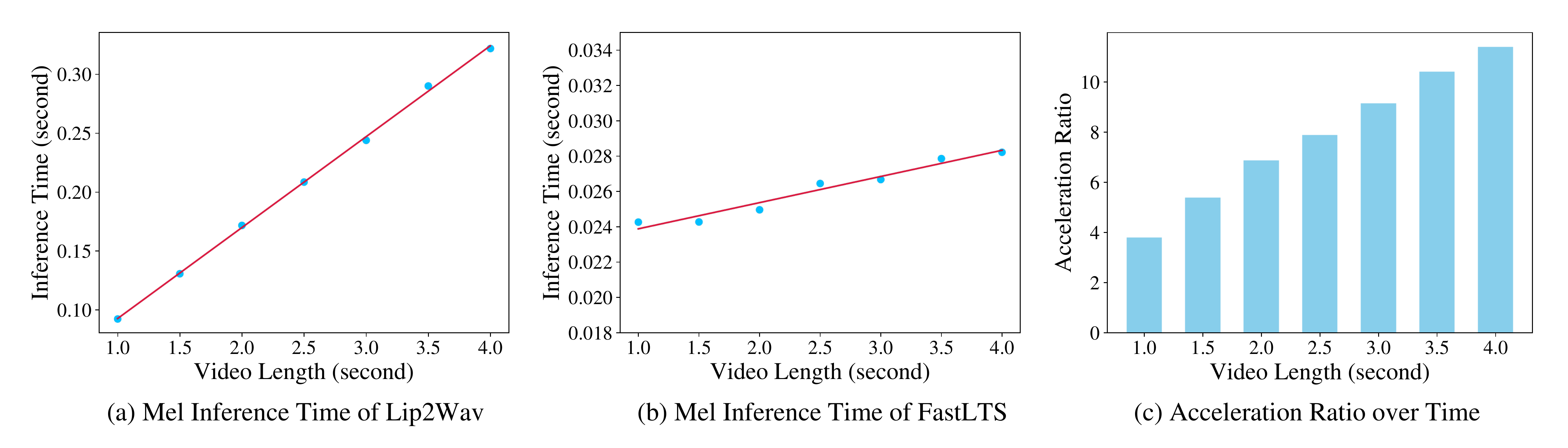}
    \caption{The Inference Speed of Mel-spectrogram of the Two Models}
    \label{fig:melspeed}
    \Description{figure description}
\end{figure*}

\begin{figure*}
    \includegraphics[width=\textwidth]{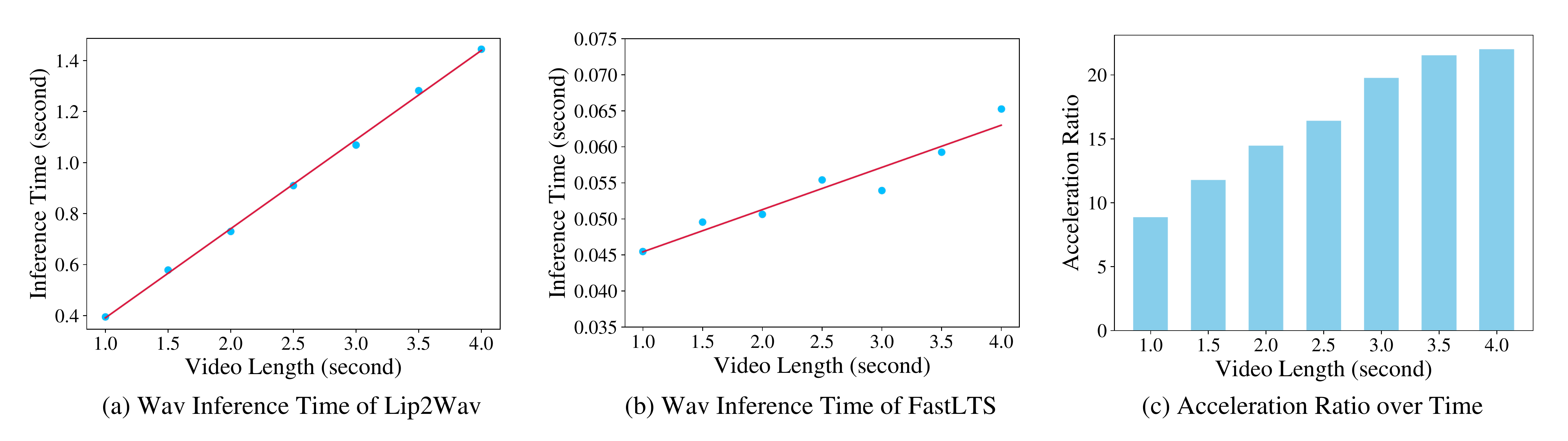}
    \caption{The Inference Speed of Waveform of the Two Models}
    \label{fig:wavspeed}
    \Description{figure description}
\end{figure*}

We also try to combine the Lip2Wav model with neural vocoders, and report the MOS in Table \ref{tab:l2wvoc_mos}. We train hifi-gan vocoders on the audios and spectrograms from the Lip2Wav dataset, and use them to generate speech waveforms from spectrograms generated by Lip2Wav. We get highly distorted speeches with a strong and harsh electronic noise. As shown in Table \ref{tab:l2wvoc_mos}, these audio samples get extremely low MOS, indicating their low quality. This matches the observation of \cite{DBLP:conf/cvpr/PrajwalMNJ20}, and proves that our model is much more than simply adding an acoustic model with a neural vocoder. Our model enables the usage of neural vocoder in unconstrained lip-to-speech synthesis, which is not possible for previous works.

For the GRID dataset, we conduct experiments on three speakers and report the MOS in Table \ref{tab:grid_mos}. It can be seen that our model achieves a comprehensive improvement on all three aspects, which is consistent with the results under unconstrained settings. We also calculate PESQ and compare our results with the scores of previous works in Table \ref{tab:gridpesq}. Despite the gap between the results of our model and the state-of-the-art method, our model has outstanding performance and ranks high on the list.

\subsection{Inference Speedup}
We measure the inference speedup by comparing the inference time of the autoregressive model and our FastLTS model on input video sequences of different lengths. We use the \textit{Chemistry Lectures} speaker in the Lip2Wav dataset to do the comparison. We conduct our experiments on a server with an Intel-Xeon E5-2678 CPU and an NVIDIA RTX-2080Ti GPU with batch size of 1.

Generally, we don't directly use the mel-spectrogram output of FastLTS, as it is only used to assist training. Yet we still compare the mel-spectorgram inference speed of the two models to eliminate the effects of using different waveform synthesis methods. The inference time and acceleration ratio are illustrated in Figure \ref{fig:melspeed}. We can see that our model generates mel-spectrograms several times faster than Lip2Wav, and the acceleration ratio increases as the input sequence gets longer. This is because our non-autoregressive model adopts parallel generation, so the inference latency doesn't increase dramatically as the input sequence gets longer, while the latency of Lip2Wav increases sharply due to its autoregressive nature. We can see that the acceleration ratio reaches $9.14\times$ when the input length is 3 seconds which is the configured input length in FastLTS and Lip2Wav.

The waveform inference time of the two models is illustrated in Figure \ref{fig:wavspeed}. We can see that the Griffin-Lim algorithm causes great inference latency in Lip2Wav, and the latency grows sharply as the mel-spectrogram sequence gets longer, while our model adopts parallel generation of audio with a GAN-based vocoder, which brings great generation speedup. The acceleration ratio of waveform synthesis reaches $19.76\times$ when the input length is 3 seconds. It is worth noting that the acceleration ratio of our model exceeds that of GlowLTS \cite{GlowLTS}. As reported in \cite{GlowLTS}, GlowLTS only achieves $5.337\times$ acceleration when the video length is 30 seconds, and is slower than the autoregressive model when the input length is shorter than 3 seconds.

\begin{table}
  \caption{Parameter Amounts of Three Different Models}
  \label{tab:size}
  \begin{tabular}{lcc}
    \toprule
    \textbf{Model}&\textbf{Parameters}&\textbf{Relative Size}\\
    \midrule
    \textit{Autoregressive Model} \\
    \midrule
    Lip2Wav & 39.87M & $1.00\times$ \\
    \midrule
    \textit{Non-autoregressive Models} \\
    \midrule
    GlowLTS & 85.92M & $2.16\times$ \\
    FastLTS(ours) & 50.09M & $1.26\times$ \\
  \bottomrule
\end{tabular}
\end{table}

\begin{table}
  \caption{CMOS Comparison in Ablation Studies}
  \label{tab:abl}
  \begin{tabular}{cccc}
    \toprule
    \textbf{Method} & \textbf{Quality} & \textbf{Intelli.} & \textbf{Natural.} \\
    \midrule
     FastLTS & 0 & 0 & 0 \\
     \midrule
    \tabincell{c}{w/o waveform generator} & -0.274 & -0.075 & -0.103 \\
    \tabincell{c}{w/o conditional module} & N/A & N/A & N/A \\
    \tabincell{c}{w/o first training stage} & N/A & N/A & N/A \\
  \bottomrule
\end{tabular}
\end{table}

\subsection{Model Size}
We compare the model size of our FastLTS with Lip2Wav and GlowLTS, and the results are shown in Table \ref{tab:size}. We can see that as the cost of realizing a non-autoregressive architecture with generative flow, GlowLTS has over twice more parameters than Lip2Wav, causing high memory usage. In contrast, our model only use 26\% more parameters to realize an end-to end non-autoregressive architecture that can generate high-quality audio with short latency. 

\section{Ablation Study}
\begin{figure}
    \includegraphics[width=\linewidth]{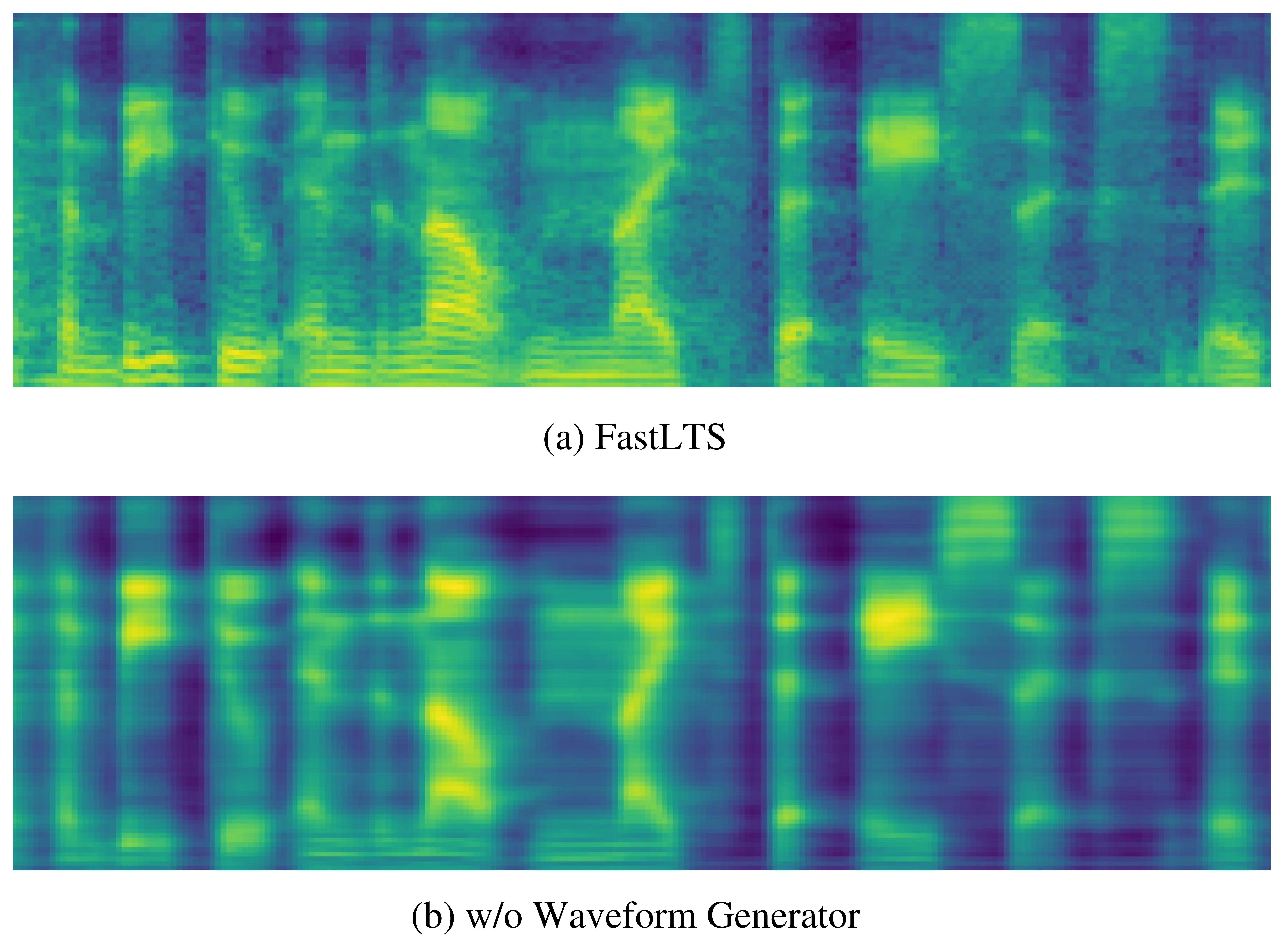}
    \caption{Comparison of mel-spectrograms for ablation studies}
    \label{fig:ablation}
    \Description{figure description}
\end{figure}

In this section, we conduct ablation experiments on FastLTS to explore the effectiveness of our model. We first remove the waveform generator and conduct CMOS evaluation on the results. The scores are shown in Table \ref{tab:abl}. We can see that the removal of the waveform generator causes drop in all three aspects, especially the audio quality, where the CMOS is -0.274. This proves that the waveform generator plays an important role in improving the quality, intelligibility and naturalness of the audio. We also visualize the mel-spectrogram samples in Figure \ref{fig:ablation}. It can be seen that the acoustic conditional module tends to generate blurry and over-smoothed spectrograms just like Lip2Wav when there is no waveform generator.

We then remove the acoustic conditional module and try to use raw waveforms to train the whole model. However, we fail to generate intelligible audios with this model. This indicates that the acoustic conditional module plays a significant role in modeling the acoustic features. By comparing the shape of the spectrograms in Figure \ref{fig:ablation}(a) and \ref{fig:ablation}(b), we can infer that acoustic conditional module is responsible for modeling the low-level structures of the audio, while the waveform generator is responsible for complementing high-dimensional details.

We also try to remove the first training stage and train the whole model in an end-to-end way from scratch. All we get, however, are blurry and indistinguishable human voices. This indicates that it is not practical to directly use the raw audio as supervisory signal to train the model, and proves the necessity of the two-stage training.

\section{Conclusions}
In this paper, we devise FastLTS, an end-to-end model for unconstrained lip-to-speech synthesis. To improve audio quality, we build an end-to-end model by using a GAN-based vocoder and applying adversarial training. To reduce inference latency, we adopt a fully parallelized architecture with a non-autoregressive decoder and the vocoder. Besides, we design an applicable transformer-based visual frontend for this task, which shows equivalent performance to 3D-CNN-based counterparts. We conduct experiments on the Lip2Wav dataset, showing significant improvements on audio quality and inference speed. We also conduct experiments on the GRID dataset, showing the adaptability of our model on small datasets.  

For future work, we will try to extend the model to multi-speaker settings. We also plan to combine our lip-to-speech synthesis model with human-face super-resolution models to tackle the problem of lip-to-speech synthesis with low-resolution videos.

\begin{acks}
This work was supported in part by the National Key R\&D Program of China (Grant No.2018AAA0100600 and No.2020YFC0832505), National Natural Science Foundation of China (Grant No.61836002 and No.62072397) and Zhejiang Natural Science Foundation (LR19F020006).
\end{acks}

\clearpage

\bibliographystyle{ACM-Reference-Format}
\bibliography{ref}


\end{document}